\documentclass[pra,twocolumn,showpacs,floats]{revtex4}
\usepackage{graphicx}
\begin{document}
\title{
Efficient implementations of the Quantum Fourier Transform:
an experimental perspective}
\author{Kavita Dorai}
\email{kavita@e3.physik.uni-dortmund.de}
\author{Dieter Suter}
\email{Dieter.Suter@physik.uni-dortmund.de}
\affiliation{Department of Physics, University of Dortmund, D-44221 Germany}
\date{\today}
\begin{abstract}
The Quantum Fourier transform (QFT) is a key ingredient in most
quantum algorithms. We have compared various spin-based quantum
computing schemes to implement the QFT from the point of view of their
actual time-costs and the accuracy of the implementation.  We focus
here on an interesting decomposition of the QFT as a product of the
non-selective Hadamard transformation followed by multiqubit gates
corresponding to square- and higher-roots of controlled-NOT gates.
This decomposition requires only $O(n)$ operations and is thus linear
in the number of qubits $n$.  The schemes were implemented on a
two-qubit NMR quantum information processor and the resultant density
matrices reconstructed using standard quantum state tomography
techniques.  Their experimental fidelities have been measured and
compared.
\end{abstract}
\pacs{03.67.Lx,76.70.-k}
\maketitle
\section{Introduction}
The Fourier transform finds
widespread application in physics
and information processing, and it comes as no
surprise that its quantum version lies at the core
of most known quantum computational algorithms.
The Quantum Fourier Transform (QFT) is analogous
to the classical Fast Fourier Transform (FFT), and
by exploiting the advantages of quantum
parallelism, can be computed exponentially faster.
However, this advantage cannot be used to speed
up data processing tasks, since the individual
Fourier transformed output amplitudes cannot
be accessed by a measurement. What the QFT {\em can}
achieve is an estimation of arbitrary quantum
phases and an extraction of the periodicity of
a function. Indeed, fast quantum algorithms
for factoring~\cite{shor,ekert-proc,jozsa-proc},
finding discrete logarithms~\cite{simon}
and the more general
algorithm for finding the stabilizer of
an Abelian group~\cite{kitaev},
rely crucially on this
property of the QFT.

Schemes to implement the QFT have been proposed
using cavity QED~\cite{scully-qed} and have been
experimentally implemented using
NMR~\cite{cory1,chinese1,cory2,chinese2}. Despite several
limitations (see for example the
points made in~\cite{jones-review, cory-review} and similar
reviews), NMR remains to date
the only existing quantum-computing
technology. However, with ideas for quantum algorithms
that work with expectation-value
quantum computers~\cite{glaser,collins} and
proposals for scalable solidstate NMR
quantum computing implementations~\cite{kane,suter,yamamoto}, it
is likely that spin-based implementations will
soon emerge as a viable
technology for quantum computing.

The key role played by the QFT in quantum
algorithms makes it an attractive candidate
for detailed investigations of its
experimental implementations.
Issues of the actual time-cost of quantum
algorithms as compared
to their ideal computational cost have
seldom been quantitatively addressed~\cite{saito,barenco}.
However, these issues are relevant and need to be tackled for
technology to keep pace with theoretical developments.
This paper seeks to compare different decompositions of
the QFT, with a view to finding the most
efficient experimental spin-based
quantum computing implementation.


\section{The QFT and its decompositions}

The basis states that we consider are product states
$$
\vert a \rangle = \vert a_{n-1} a_{n-2}...a_0 \rangle =
\vert a_{n-1} \rangle_{n-1} \otimes ...\vert a_1 \rangle_1 \otimes
\vert a_0 \rangle_0 ,
$$
which can be represented by binary integers
$$
a = \sum_{j=0}^{n-1} a_j 2^j, a_j \in \{0,1\} .
$$
q = $2^n$ is the dimension of the Hilbert space and $n$ the number of qubits.

In this basis, the QFT can be represented as a unitary operator $\mathcal{F}$,
which transforms the basis states $\vert a \rangle$ into
\begin{equation}
\mathcal{F} \vert a \rangle =
\frac{1}{\sqrt{q}} \sum_{c=0}^{q-1} e^{2 \pi i a c/q} \vert c \rangle .
\label{qft-eqn1}
\end{equation}
The states $\vert c \rangle$ have the same form as $\vert a \rangle$.

When applied to an arbitrary state
$ \vert \psi \rangle = \sum_a A_a \vert a \rangle$, the QFT yields
\begin{equation}
\mathcal{F} \vert \psi \rangle
\equiv \mathcal{F} \sum_{a=0}^{q-1}
A_a \vert a \rangle  \longrightarrow
\sum_{c=0}^{q-1} C_c \vert c \rangle
\end{equation}
where the coefficients $C_c$ are the
discrete Fourier transform of the input coefficients $A_a$.

The basis transformation of Eqn.~\ref{qft-eqn1} can be written in
terms of individual qubits as
\begin{equation}
\mathcal{F} \vert a \rangle = \prod_{j=0}^{n-1}
\otimes \vert p(\phi_j) \rangle_j
\label{qft-eqn2}
\end{equation}
where each qubit is in a state
$\vert p(\phi_j) \rangle = (\vert 0 \rangle + e^{i 2 \pi \phi_j}
\vert 1 \rangle)/\sqrt{2}$.
The phases are determined by
$$
\phi_j = \sum_{k=0}^{n-1-j} a_k 2^{j+k-n}
$$

Equation \ref{qft-eqn2} serves as the basis for implementing the QFT
by one- and two-qubit operations.
One implementation~\cite{barenco,coppersmith,griffiths} uses
single-qubit Hadamard rotations gates $H_j$ and two-qubit
controlled-phase gate $B_{j,k}$ that act on the qubits $j$ and $k$
and are given by
$$
B_{j,k} = \left(\begin{array}{cccc}
1&0&0&0 \\
0&1&0&0 \\
0&0&1&0 \\
0&0&0&e^{i \theta_{jk}}
\end{array}
\right)
$$
where $\theta_{jk} = \pi 2^{j-k}$ is a conditional phase shift applied only if
both qubits are in the state $\vert 1 \rangle$.
In terms of these gates, the quantum circuit for $n$ qubits is
\begin{eqnarray}
\mathcal{QFT}_n &=&
(H_1B_{1,2}...B_{1,n})
(H_2B_{2,3}...B_{2,n}) 
...\nonumber \\
&&...(H_{n-1}B_{n-1,n})(H_n)
\label{coppersmith}
\end{eqnarray}
with the sequence of operations being performed from
right to left. With this implementation, the bit values of the result appear in
reversed order. If a sequence reversal is required, this can be achieved by a
sequence of SWAP operations on pairs of qubits.

We shall denote this decomposition of the QFT, as ``serial''.
For $n$ qubits, it requires a total of $n$
$H_j$ gates, $n(n-1)/2$ $B_{j,k}$ gates and
$n/2$ SWAP operations, leading to a computational
complexity of $O(n^2)$.

Individual Hadamard operations are qubit-selective and
hence costlier than a total Hadamard operator
that is applied on all qubits simultaneously.
It would be thus desirable to
have a decomposition of the QFT that involves a
non-selective Hadamard transformation~\cite{cory2}.
\begin{figure}
\includegraphics[width=8cm]{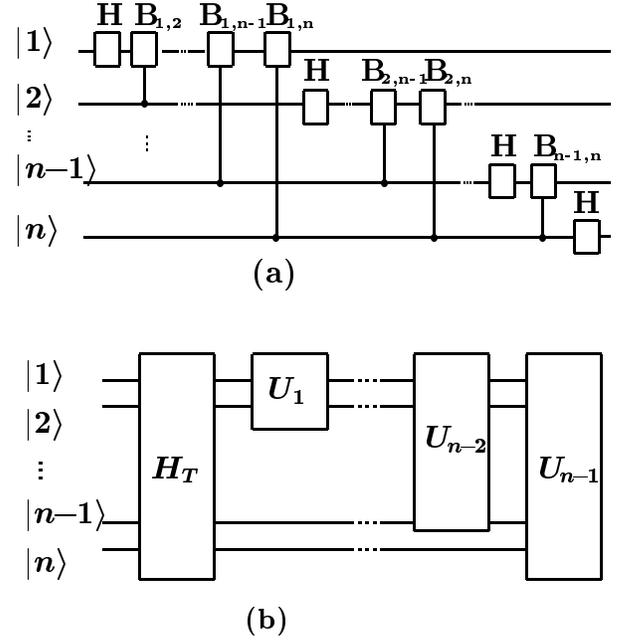}
\caption{\label{ckt}
(a) Quantum circuit for the serial implementation of the QFT using
qubit-selective Hadamard rotations and two-qubit controlled-phase gates.
(b) Circuit for the parallel
implementation of the QFT using
a non-selective Hadamard rotation on
all qubits, and multiqubit gates. The
readout of the QFT is performed in reverse
order on the qubits, achieved by SWAP operations
(not shown in the circuits).}
\end{figure}

A more useful (for NMR) decomposition of the QFT can be obtained by 
noting that the Hadamard operator is self-inverse
and that a Hadamard rotation of the controlled-phase gate can 
be decomposed as
a root of a controlled-NOT gate ,
\begin{eqnarray*}
H_k B_{jk} H_k^{-1} =
e^{i \pi/2^{k-j+1}} (U_{CNOT})^{1/2^{k-j}}_{j,k} 
\end{eqnarray*}
where $j$ is the control and $k$ the target qubit.
The global phase factor does not influence measurement results and is 
henceforth ignored.
Further, using
$$
[H_i, B_{j,k}]=0, i \ne j,k ,
$$
the sequence of operations in Eqn.~(\ref{coppersmith}) can be modified to
\begin{equation}
\mathcal{QFT}_n = [H_T][U_1U_2...U_{n-2} U_{n-1}]
\label{multiqubit}
\end{equation}
where
$$
H_T = H_1 H_2 H_3...H_{n-1}H_n 
$$
is the total non-selective Hadamard operator on all qubits, i.e. a 
single $\pi/2$ radio frequency pulse.

The $U$ gates in Eqn.~(\ref{multiqubit}) are
\begin{eqnarray}
U_{1} &=& H_2 B_{1,2} H_2^{-1} \nonumber \\
         &=& (U_{CNOT})^{1/2}_{1,2} \nonumber \\
U_{n-2} &=& H_{n-1} (B_{1,n-1}B_{2,n-1}...
B_{n-2,n-1}) H_{n-1}^{-1} \nonumber \\
&=& (U_{CNOT})^{1/2^{n-2}}_{1,n-1}
(U_{CNOT})^{1/2^{n-3}}_{2,n-1}..(U_{CNOT})^{1/2}_{n-2,n-1} \nonumber \\
U_{n-1} &=& H_n 
(B_{1,n} B_{2,n}...B_{n-1,n})
H_{n}^{-1}\nonumber \\
&=&
(U_{CNOT})^{1/2^{n-1}}_{1,n}
(U_{CNOT})^{1/2^{n-2}}_{2,n}..(U_{CNOT})^{1/2}_{n-1,n}\nonumber \\
\end{eqnarray}
They correspond to single spin rotations conditioned on the status of 
all the other spins involved in the operation.
Since they are single spin rotations, they can be implemented as single 
radio frequency pulses.
The conditioning on the state of the other spins is achieved by 
making them selective on specific transitions.
As an example, for a system of three
qubits, the operation $U_{n-1}$ in this case is given by
a fourth-root of a controlled-NOT gate 
on qubits one and three, followed by 
a square-root of a controlled-NOT gate 
on qubits two and three, with the 
third qubit being the target qubit in both cases. These gates thus
involve three transitions of the third qubit:
$100 \rightarrow 101, 110 \rightarrow 111$, 
and $ 010 \rightarrow 011$. 
Since these are unconnected transitions, rotations in the
subspace of these transitions can be achieved simultaneously.
Many pulsed irradiation schemes for such precise selective
excitation exist in NMR, mostly involving shaping the
excitation profile of the rf waveform~\cite{cory-newsel}.
The entire QFT operation in this decomposition therefore 
reduces to a sequence of $n$ radio 
frequency pulses.
It scales linearly with the number of qubits, and we will denote 
it as the ``parallel'' implementation.

\section{Time-Cost of the QFT}
The main issue in the experimental implementation
of quantum algorithms is not the number of logical
operations per se, but the actual time-cost of
each logical operation/quantum gate.
The $U$ transformations in
Eqn.~(\ref{multiqubit}) are no longer
two-qubit phase-shift gates but correspond to
square- and higher-roots of controlled-NOT operations on
specific qubits.
They can be implemented experimentally using
multiqubit gates that perform
manipulations on qubits simultaneously.
Since the NMR Hamiltonian has terms connecting
multiple pairs of qubits, such multiqubit gates
can be directly implemented.
The quantum circuits for both serial and parallel
implementations of the QFT are shown in
Figure~\ref{ckt}.

The most expensive operation in the serial
implementation of the QFT is the controlled-phase shift
gate $B_{jk}$.
The ideal time-cost is computed
assuming all gates take the same amount of
time. However experimentally, the
controlled-phase shift gate
requires a time $\tau_{jk}$ proportional to
the desired phase rotation angle
$\theta_{jk}$ (related to the ``distance''
$(k-j)$ of the qubits), and inversely
proportional to the interaction $J_{jk}$
between the qubits. The magnitude of the
interaction $J_{jk}$ and hence the time
cost depends on the specific
experimental quantum computing technology
under consideration. For liquid-state
NMR, $J_{jk}$ is the electron-mediated
scalar coupling between the qubits. For
our calculations, we assume the $J_{jk}$'s
to be of the same order of magnitude for
all qubits, represented by a universal
constant coupling $J$.
The actual time cost of the serial
decomposition of the QFT, involving only
one-qubit Hadamards and the two-qubit
phase-controlled gate is
\begin{eqnarray}
T_{ser} = n\delta+
\sum_{j=0}^{n-1} \sum_{k=j+1}^{n}
\tau_{jk}
&=&
n\delta + \kappa
\sum_{j=0}^{n-1} \sum_{k=j+1}^{n}
2^{j-k}   \nonumber \\
&=&
n\delta + \kappa (n - 1 + 2^{-n}) \nonumber \\
&&\approx O(n)
\end{eqnarray}
where $\delta$ is the time-cost of each
single-qubit Hadamard rotation and $\kappa = \pi/J$.
Using multiqubit
gates in the parallel implementation
of the QFT reduces the actual time-cost of
the algorithm. Quite apart from the saving
obtained by using a non-selective
Hadamard transformation in the beginning
on all the qubits, each $U$ gate can be
thought of as having components from
one or more $B_{jk}$ gates.
The actual time-cost of the parallel QFT is
\begin{equation}
T_{par} =
\kappa \sum_{j=0}^{n-1} \sum_{k=j+1}^{n}
\tau_{jk}
\label{partial-parallel}
\end{equation}
Since for multiqubit gates, the system
evolves under more than one coupling
period simultaneously, only the largest of
these need be counted for contribution to
the time-cost and the inner sum in
Eqn.~(\ref{partial-parallel}) vanishes to give
\begin{equation}
T_{par} = \kappa \sum_{j=0}^{n-1} 2^{-1}
= \kappa n/2 \approx O(n)
\end{equation}
The analysis does not include the degradation
of each gate due to decoherence nor does it
take into account the SWAP operations, since the
latter can in most cases be avoided by a
relabeling of qubits. Implementing the
Approximate QFT~\cite{barenco} would require
fewer controlled-phase gates but
would correspondingly reduce the accuracy.
\section{NMR Implementations of the QFT}
Experiments were performed on a de-gassed,
flame-sealed sample of ${}^{13}$C-labeled
chloroform, with ${}^{13}$C and ${}^{1}$H
as the two qubits and a coupling constant of
$J_{12}\approx 215$Hz.
Qubit-selective $90$ degree
pulses are of the order of $10 \mu s$.
The unitary transformations required for the
parallel decomposition of the QFT can be
implemented either by transition-selective
pulses or by J-coupling intervals sandwiched
between qubit-selective pulses.
A low-power rectangular pulse of length
$6.5 ms$ was used to selectively excite
individual transitions for the selective
implementation of the QFT.
For heteronuclear systems, RF pulses are
applied on two different channels, leading
to a reduction in the duration of
selective pulses.

Each version of the QFT was implemented on
a temporally averaged
pseudopure state~\cite{temp-avg},
obtained from the thermal equilibrium
ensemble as the sum of three experiments
\begin{eqnarray*}
{\bf E} \,\,\,\,\,\,\,\, \mbox{(do-nothing operation)} \nonumber \\
\{90_x\}^C \frac{1}{2 J_{12}}
\{90_y\}^{C}\{90_x\}^{H} \frac{1}{2 J_{12}}
\{90_y\}^H \nonumber \\
\{90_x\}^H \frac{1}{2 J_{12}}
\{90_x\}^{C}\{90_y\}^{H} \frac{1}{2 J_{12}}
\{90_y\}^C \nonumber \\
\end{eqnarray*}
The details of the pulse sequences used to
implement the serial, parallel and
the selective-pulse (parallel) decompositions
of the QFT, are given in Table~\ref{pulse-table}.
The final SWAP operation was not executed; instead
the readout in the reverse order was achieved by
``relabeling'' the qubits at the end of each
experiment.

\begin{figure}
\includegraphics[width=8cm]{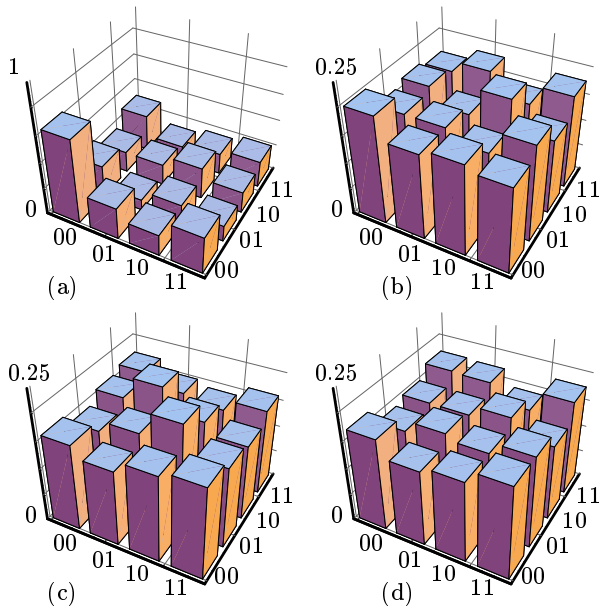}
\caption{\label{tomo-fig} The
experimental deviation density matrices
(a) for the pseudopure state $\vert 0 0 \rangle$
and for the states obtained after applying the
QFT to the pseudopure state using the
pulse sequence for (b) the selective implementation
(c) the parallel implementation and (d) the
serial implementation of the QFT. The rows
are labelled in the standard computational basis.}
\end{figure}

The results of all three implementations of the
QFT are shown in Figure~\ref{tomo-fig}, using
three-dimensional bar graphs to represent
components of the final density matrix. Since
only single-quantum terms are observable in
NMR, it is necessary to perform a series of
experiments that rotate unobservable terms
into observable ones, in order to sample the
entire density matrix. The density matrix
after each implementation of the QFT was
reconstructed by standard quantum state
tomography procedures, using a set of nine
experiments and qubit-selective
readouts~\cite{tomo,tomo-chinese}.

The precision of the QFT implementation
can be estimated by measuring its ``fidelity'',
defined for mixed density matrices (such
as the ones encountered in NMR)~\cite{cory1}
$$
F = \frac{Tr(\rho_{th} \rho_{exp})}
{\sqrt{Tr(\rho^2_{th})}\sqrt{Tr(\rho^2_{exp})}}
\sqrt{\frac{Tr(\rho^2_{exp})}
{Tr(\rho^2_{init})}}
$$
The first term in the expression measures the
correlation between the experimental
deviation density matrix $\rho_{exp}$ and
the theoretical deviation density matrix
$\rho_{th}$ (obtained by ``applying'' the
unitary operator corresponding to the
ideal QFT transformation, to the
initial density matrix $\rho_{init}$).
The second term is the weighting factor
to take into account the overall signal
loss due to decoherence during the experiment.

The fidelities measured for the serial,
parallel and selective-pulse versions of
the QFT are $79\%$, $80\%$ and $85\%$ respectively.
The reduction in fidelity is mainly due to
imperfections in pulse calibration as well
as system decoherence. It is not surprising that
the serial and parallel versions are equally
accurate for the case of two qubits. The savings
in time and the increase in accuracy of the
parallel QFT will be realised only for a larger
number of qubits. The better performance of the
selective scheme is due to the fact that such
a direct implementation of the square-root of
the controlled-NOT gate does not require refocusing
schemes~\cite{freeman,kd}. However, in systems with a larger number of
qubits such selective pulse schemes might not be feasible,
the major stumbling
blocks in such cases being decoherence during
the pulses and the overlap of transitions in crowded spectra.
\section{Other spin-based architectures}
Recently, several approaches have been suggested
for the design of solid-state spin-based quantum
computers. Kane's proposal~\cite{kane} using
single donor spins in Si, addresses the problem
of scalability but has the disadvantages
inherent in single-spin measurements.
Ladd et.~al.'s solid-state NMR quantum computing
device on the other hand, is made entirely
of silicon, with the qubits being spin-1/2
nuclei located in isolated
atomic chains~\cite{yamamoto}.
Suter et.~al.~\cite{suter} proposed an
alternative architecture with each logical
qubit being represented by two physical
qubits - an active electron
spin to manipulate quantum
information and a passive nuclear
spin to store information.
A logical qubit is addressed using magnetic
field gradients and SWAP gates, realised
as a cascade of three transition-selective
pulses,  are used to
convert between active and passive states.
A basic two-qubit gate relies on the
dipolar interaction between electron spins
and requires four additional SWAP gates, two
to switch between active and passive states
and two back-SWAPS to switch off the interaction
between the neighbouring qubits. The hyperfine
interaction is of the order of a few MHz and the
electron dipolar interaction strength is around
10-50 MHz. An estimate of the actual time-cost
of the QFT for such a solid-state spin quantum computer
yields
\begin{eqnarray}
T_{ser} &=& n\delta+
\sum_{j=0}^{n-1} \sum_{k=j+1}^{n}
(\tau_{jk}+2*\tau_{\tiny SWAP}) \nonumber \\
&=&
n\delta +  2n\Delta +\kappa
\sum_{j=0}^{n-1} \sum_{k=j+1}^{n}
2^{j-k}   \nonumber \\
&=&
n\delta + 2n\Delta + \kappa (n - 1 + 2^{-n})
\approx O(n)
\end{eqnarray}
where $\delta$ is the time-cost of each
single-qubit Hadamard rotation, $\kappa = \pi/d$,
$d$ is the strength of the dipolar interaction
and $\Delta$ is the time unit of one SWAP gate.
Since the gate times for this
implementation are very fast, a greater number
of logical operations compared to
liquid-state NMR computers, can be performed within the
system decoherence limit. These solid-state
proposals are also scalable to a very large
number of qubits.

In conclusion, we have estimated the realistic
time-costs of different decompositions
of the QFT for liquid and solid-state NMR quantum
computers and have measured the accuracy of
the implementations experimentally using liquid-state NMR.
While all quantum computation can be implemented
using the two-qubit universal controlled-NOT gate
and one-qubit rotations, the number of these
basic operations increases exponentially with
the number of qubits. It has been suggested that
for some specific QC purposes, using more
complicated multiqubit gates might be computationally
more efficient~\cite{multiqubit1,multiqubit2}.
The parallel
implementation of the QFT suggested by Cory et.~al.,
using multiqubit gates, performs better than the serial
implementation.  The actual experimental time-costs
can be improved upon
using innovative techniques like multiqubit
gates, creative refocusing
schemes~\cite{refocus} and time-optimal
gates designed using control
theory~\cite{glaser-control,chinese-control}.

\begin{table*}
\caption{
NMR pulse schemes for different implementations
the QFT on $n = 2$ and 3 qubits.
Superscript $r \rightarrow s$ indicates
a selective RF pulse on the transition $r \rightarrow s$.
Superscripts $r$ and $r,s$ indicate spin-selective
pulses on the spins $r$ and $r,s$ respectively.
Subscript $z$ indicates a composite-z pulse
which can be expanded as a sandwich of rf pulses
$\{\theta_z\} \equiv \{90_x\} \{\theta_y\}\{90_x\}$.}
\begin{ruledtabular}
\begin{tabular}{ccc}
Implementation  & $n=2$ & $n=3$ \\
\hline
Serial QFT &
$\{90_y\}^1 \{180_x\}^1
\frac{1}{4J_{12}} \{90_y\}^{1,2}
\{45_x\}^{1,2}$ &
$\{45_y\}^3 \{180_x\}^3 \{45_{-y}\}^3
\{180_x\}^3 \frac{1}{4 J_{23}}$\\
& $ \{90_{-y}\}^{1,2}
\{90_y\}^2 \{180_x\}^2$ &
$\{180_{-x}\}^3 \{90_y\}^{2,3}
\{45_x\}^{2,3} \{90_{-y}\}^{2,3}
\{45_y\}^2$ \\ 
&& $\{180_x\}^2 \{45_{-y}\}^2
\{180_x\}^1 \frac{1}{8 J_{13}}
\{180_{-x}\}^1 $ \\
&& $\{90_y\}^{1,3}
\{22.5_x\}^{1,3} \{90_{-y}\}^{1,3}
\{180_x\}^2 \frac{1}{4 J_{12}}$ \\
&& $\{180_x\}^2
\{90_y\}^{1,2} \{45_x\}^{1,2}
\{90_{-y}\}^{1,2} \{45_y\}^1$ \\
&& $\{180_x\}^1 \{45_{-y}\}^1$ \\
\hline
Parallel QFT &

$\{90_y\}^{1,2} \{180_x\}^{1,2} \{90_y\}^2
\{180_x\}^{1,2}
\frac{1}{4J_{12}}$&

$\{45_y\}^{1,2,3} \{180_x\}^{1,2,3}
\{45_{-y}\}^{1,2,3} \{90_{-y}\}^2
\frac{1}{4 J_{23}}$\\ 

&&$\{90_y\}^2 \{45_x\}^2
\{180_x\}^1
\{90_y\}^{1,2}
\{45_x\}^{1,2}$ \\
&& $\{90_{-y}\}^{1} \{180_x\}^2
 \{90_{-y}\}^1 \{\frac{1}{4 J_{12}},
\frac{1}{8 J_{13}}\} \{90_y\}^1$\\
&& $\{67.5_x\}^1 \{135_{-z}\}^3 \{90_{-z}\}^2$ \\

\hline
Selective-pulse &$\{90_y\}^{1,2}
\{180_x\}^{1,2} \{90_x\}^{3 \rightarrow 4}
\{45_z\}^1$  &
$\{90_y\}^{1,2,3} \{180_x\}^{1,2,3}
\{90_x\}^{6 \rightarrow 8} \{90_x\}^{5 \rightarrow 7}$ \\
(parallel QFT) & &
$\{180_x\}^{7 \rightarrow 8} \{90_x\}^{5 \rightarrow 6}
\{90_x\}^{3 \rightarrow 4}
\{67.5_z\}^1 \{45_z\}^2$ \\
\end{tabular}
\label{pulse-table}
\end{ruledtabular}
\end{table*}
\begin{acknowledgments}
All experiments were performed on a Bruker DRX-500 MHz
NMR spectrometer located in the Chemistry Dept.,Dortmund
University. K.D.
is supported by a
post-doctoral fellowship from the
DFG-funded
Graduiertenkolleg
GK-726/1-02.
\end{acknowledgments}


\begin{thebibliography}{99}
\bibitem{shor}
P.~W.~Shor, SIAM J.~Comput., {\bf 26}, 1484 (1997).
\bibitem{ekert-proc}
R.~Cleve, A.~Ekert, C.~Macchiavello,
and M.~Mosca,
Proc.~R.~Soc.~A, {\bf 454}, 339 (1998).
\bibitem{jozsa-proc}
R.~Jozsa, Proc.~R.~Soc.~A, {\bf 454}, 323 (1998).
\bibitem{simon}
D.~R.~Simon, SIAM J.~Comput., {\bf 26}, 1474 (1997).
\bibitem{kitaev}
A.~Y.~Kitaev, quant-ph/9511026 (1995).
\bibitem{scully-qed}
M.~O.~Scully and M.~S.~Zubairy,
Phys.~Rev.~A, {\bf 65}, 052324 (2002).
\bibitem{cory1}
Y.~S.~Weinstein, S.~Lloyd, and D.~G.~Cory,
Phys.~Rev.~Lett., {\bf 86}, 1889 (2001).
\bibitem{chinese1}
L.~Fu, J.~Luo, L.~Xiao, and X.~Zheng,
quantph/9905083 (1999).
\bibitem{cory2}
M.~D.~Price, T.~F.~Havel, and D.~G.~Cory,
New J.~Phys., {\bf 2}, 101 (2000).
\bibitem{chinese2}
X.~Peng, X.~Zhu, X.~Fang,
M.~Feng, X.~Yang, M.~Liu, and
K.~Gao,  quant-ph/0202010 (2002).
\bibitem{jones-review}
J.~A.~Jones, Fortschr.~Phys., {\bf 48},
909 (2000).
\bibitem{cory-review}
D.~G.~Cory, R.~Laflamme, E.~Knill et.~al.,
Fortschr.~Phy., {\bf 48}, 875 (2000).
\bibitem{glaser}
J.~M.~Myers, A.~F.~Fahmy, S.~J.~Glaser, and
R.~Marx, Phys.~Rev.~A, {\bf 63}, 032302 (2001).
\bibitem{collins}
D.~Collins, Phys.~Rev.~A, {\bf 65}, 052321 (2002).
\bibitem{kane}
B.~E.~Kane, Fortschr.~Phys., {\bf 48}, 1023 (2000).
\bibitem{yamamoto}
T.~D.~Ladd, J.~R.~Goldman, F.~Yamaguchi, and
Y.~Yamamoto, Phys.~Rev.~Lett.,
{\bf 89}, 017901 (2002).
\bibitem{suter}
D.~Suter and K.~Lim, Phys.~Rev.~A, {\bf 65},
052309 (2002).
\bibitem{saito}
A.~Saito, K.~Kioi, Y.~Akagi, N.~Hashizume,
and K.~Ohta, quant-ph/0001113 (2000).
\bibitem{barenco}
A.~Barenco, A.~Ekert,
K-A.~Suominen, and P.~Torma, Phys.~Rev.~A, {\bf 54},
139 (1996).
\bibitem{coppersmith}
D.~Coppersmith, IBM Res.~Rep., {\bf RC 19642} (1994).
\bibitem{griffiths}
R.~B.~Griffiths and C-S.~Niu, Phy.~Rev.~Lett.,
{\bf 76}, 3228 (1996).
\bibitem{cory-newsel}
E.~M.~Fortunato, M.~A.~Pravia, N.~Boulant,
G.~Teklemariam, T.~F.~Havel, and D.~G.~Cory,
J.~Chem.~Phys., {\bf 116}, 7599 (2002).
\bibitem{temp-avg}
E.~Knill, I.~Chuang, and R.~Laflamme,
Phys.~Rev.~A, {\bf 57}, 3348 (1998).
\bibitem{tomo}
I.~L.~Chuang, N.~Gershenfeld, M.~G.~Kubinec,
and D.~W.~Leung, Proc.~R.~Soc.~A,
{\bf 454}, 447 (1998).
\bibitem{tomo-chinese}
G.~L.~Long, H.~Y.~Yan, and Y.~Sun,
J.~Opt.~B, {\bf 3}, 376 (2001).
\bibitem{freeman}
N.~Linden, H.~Barjat, and R.~Freeman,
Chem.~Phys.~Lett., {\bf 296}, 61 (1998).
\bibitem{kd}
K.~Dorai, Arvind, and A.~Kumar,
Phys.~Rev.~A, {\bf 61}, 042306 (2000).
\bibitem{multiqubit1}
M.~D.~Price, S.~S~Somaroo, A.~E.~Dunlop et.~al.,
Phys.~Rev.~A, {\bf 60}, 2777 (1999).
\bibitem{multiqubit2}
J.~F.~Du, M.~J.~Shi, J.~H.~Wu et.~al.,
Phys.~Rev.~A, {\bf 63}, 042302 (2001).
\bibitem{refocus}
J.~A.~Jones, and E.~Knill, J.~Magn.~Res.,
{\bf 141}, 322 (1999).
\bibitem{glaser-control}
N.~Khaneja, S.~J.~Glaser, and R.~Brockett,
Phys.~Rev.~A, {\bf 65}, 032301 (2002).
\bibitem{chinese-control}
R.~Wu, C.~Li and Y.~Wang, Phys.~Lett.~A,
{\bf 295}, 20 (2002).
\end{thebibliography}
\end{document}